\documentclass[aps,prl,onecolumn,groupedaddress,showpacs,showkeys]{revtex4}
\bibliography{comment2.bbl}
\usepackage{graphicx}

\begin{document}

\newcommand{\be}{\begin{equation}}
\newcommand{\ee}{\end{equation}}

\title{A robust superconducting setup to probe the thermal Casimir effect}

\date{\today}

\author{Giuseppe Bimonte}
\affiliation{Dipartimento di Scienze
Fisiche, Universit{\`a} di Napoli Federico II, Complesso Universitario
MSA, Via Cintia, I-80126 Napoli, Italy}
\affiliation{INFN Sezione di
Napoli, I-80126 Napoli, Italy }

\begin{abstract}

We describe a  superconducting Casimir apparatus inspired by a recently proposed setup involving magnetic surfaces [G. Bimonte, Phys. Rev. Lett. {\bf 112}, 240401 (2014)]. The present setup consists of a superconducting Nb sphere and a flat gold plate including in its interior a superconducting Nb strip.
The experimental scheme involves a differential measurement  of the Casimir force at a point of the gold plate above the Nb strip and away from from it.    We show that similar to the previous setup, 
the superconducting system considered here implies widely different modulations of the Casimir force,  depending on whether the thermal force is computed using  the Drude or the plasma model, thus paving the way   to an unambiguous discrimination between these alternative  prescriptions. 

\end{abstract}

\pacs{05.40.-a, 42.50.Lc,74.45.+c}
\maketitle

\section{Introduction}

 The last 20 years withnessed a resurgence of interest in the Casimir effect \cite{Casimir48}, triggered by a series of new precision experiments. Impressive progress has been made both in theoretical modelling, as well as on the experimental side.  On the former front, powerful scattering methods have been developed \cite{lambr,rahi} by which it is in principle possible to compute the Casimir force between two bodies of arbitrary shapes and materials. On the other hand many new experiments have been performed to probe the Casimir effect for non-planar microstructured surfaces \cite{chen,chan,bao,chiu,bani,deccanat}, and for many diverse materials (other than conductors) like for instance  semiconductors \cite{semic}, liquids \cite{liq}, superconductors \cite{superc} and magnetic materials \cite{bani2}.      

The overall picture is rather satisfying, as it has become abundantly clear that we have achieved a good theoretical understanding of the main factors that      
significantly affect the   Casimir force between real material surfaces,  like their geometrical and optical features,  
as well as possible imperfections of the surfaces like  roughness, absorbed impurities,  patch potentials  and/or stray charges on the surfaces. For a comprehensive review see \cite{book1,book2}.
 
In this landscape of successes, however, there still is a dark spot, since an important problem is still waiting for a satisfactory theoretical explanation, as well as for a clear experimental resolution.  
The problem is about the influence of the temperature on the magnitude of the Casimir force, in the experimentally important case of metallic surfaces. It is useful to briefly review the matter. 
For finite temperatures, the modes of the em field propagating
in the region of space between two surfaces get populated according to the Bose-Einstein distribution, thus modifying the free-energy of the system. 
The resulting correction to the Casimir force is quite easy to compute  for two perfectly conducting parallel plates. One finds that for separations $a$ smaller than the thermal length $\lambda_T=
\hbar c/(2 \pi k_B T)$ ($\lambda_T=1.2 \mu$m at room temperature) the thermal correction  scales like $(a/\lambda_T)^4$ and is negligibly small, while for  $a \gg \lambda_T$  it  is {\it linear} in the temperature
$F_T^{({\rm id };\infty)}=k_B T \zeta(3)/(4 \pi a^3)$ and it represents the dominant contribution to the Casimir force. To the costernation of the experts, it was realized in 2000 \cite{sernelius} that the above simple findings are in sharp contrast with the results of detailed computations taking into full account the finite conductivity of real metals.
It was found that  the   thermal correction  scales {\it linearly} in the temperature also  for small separations, and  it was orders of magnitude larger than the one for an ideal metal   (even though  it still represented a small correction to the total force, of a few percent for separations around 400 nm at room temperature). Moreover its limiting value for large separations   was only one-half the perfect-conductor limit $F_T^{({\rm id };\infty)}$!  
A further perplexing feature of the real metal result was  that the thermal force   could not be reconciled with the perfect conductor limit by taking the conductivity to infinity. 
It was also found that the finite conductivity result, while consistent at large separations with the Bohr-van Leeuwen theorem of classical statistical physics \cite{Martin,bimo2},  appeared to violate Nernst third law of thermodynamics in the {\it idealized} limit of two plates with a perfect crystal structure \cite{bezerra,bezerra2}. No thermodynamic inconsistences arise however if account is taken of the finite residual resitance of real metals at low temperatures \cite{brevik2,brevik3,brevik4}. 

A thorough investigation quickly revealed  the mathematical reason for the large $T$-linear thermal correction. One has to recall that the famous  Lifshitz formula \cite{lifs} provides a representation of the 
Casimir free energy for two parallel slabs in the form of a sum over so-called (imaginary) Matsubara frequencies  $\xi_l=2 \pi l k_B T/\hbar$, $l=0,1,2,\dots$ of a series whose terms involve the product of the Fresnel reflection coefficients of the slabs. 
It was found that the large thermal correction found in \cite{sernelius} originated from vanishing of the contribution to the Matsubara sum of the  zero-frequency transverse-electrical TE mode (TE $\omega=0$). The absence of this mode is due  to the fact that the Fresnel reflection coefficient of a metallic plate is zero for TE $\omega=0$ modes. This is very plausible since TE $\omega=0$ modes are nothing but static magnetic fields, and ohmic conductors are transparent to magnetic fields.     
The physical explanation  of the $T$-linear (repulsive) correction was elucidated by studying the  real-frequency spectrum of the thermal Casimir force. This analysis  revealed  
that the the $T$-linear thermal force originated from the interaction among the Johnson currents that exist inside conductors at finite temperature \cite{john,carsten}.

The thermodynamical inconsistencies brought about by the Drude prescription motivated some rersearchers to abandon this prescription.  It was found indeed that
a thermal correction in agreement with  Nernst theorem,  proportional to $T^3$  at small distances,
could be obtained \cite{bezerra,bezerra2} if the contribution   to the Casimir force from the  TE $\omega=0$  was calculated in accordance with the plasma model of infrared optics $\epsilon_{\rm plasma}=1-\omega_p^2/\omega^2$, with $\omega_p$ the plasma frequency  (this has since been dubbed as {\it plasma prescription}). Several small-distance Casimir experiments have been interpreted as being in accordance with the plasma prescription, and to rule out the Drude prescription (for a review of several of these experiments see \cite{critiz}). 

The view is widely shared that there is a need for experiments probing the thermal Casimir force. So far, there is only one experiment by the Yale group \cite{lamorth} which claims to have observed the thermal force between a large sphere and a plate both covered with gold, in the wide range of separations form 0.7 to 7.3 $\mu$m. The results have been interpreted by the authors as being in accordance with the  Drude prescription. This experiment has been criticized  \cite{critiz}, because the thermal Casimir force was obtained only after subtracting from the total measured force a much larger  force of uncertain nature, that the authors attributed  to electrostatic forces originating from large  patches on the surfaces.
The subtraction was perfomed by making a fit of the total observed force, based on a two-parameter model of the electrostatic force,   and not by a direct and independent measurement
of the electrostatic force, as it would have been desirable.

As of now there is no reported observation of the thermal Casimir force at separations less than 0.7 $\mu$m, which is the range in which the Casimir force has been measured most accurately. The reason is that, as note before,  the   thermal Casimir force  represents a small correction at sub micron separations,  and therefore  tight experimental demands  must be fulfilled to observe it: these include a very accurate electrostatic calibration of the apparatus  \cite{onof,onof2,onof3,decca,iannuzzi2,dalvit2},  a detailed knowledge of the optical properties of the involved surfaces  over a wide frequency range \cite{svetovoy, svetovoy2, bimonteKK} and a precise determination of the separation between the two surfaces. These factors represent sources of systematic errors that are difficult to quantify, thus complicating the interpretation  of the experiments and often stimulating  controversies. 

In order to circumvent the drawbacks of current Casimir experiments aiming at the observation of the thermal force, the author proposed recently an alternative experimental scheme \cite{hide}.  The setup described in \cite{hide} consists of two aligned sinusoidally corrugated Ni surfaces, one of which is “hidden” by a thin opaque layer of gold with a flat exposed surface. The gold over-layer, which acts as a low-pass filter,  is the key ingredient of the setup. The scheme of measurement is based on the observation of the phase-dependent modulation of the Casimir force, as one of the two corrugations is laterally displaced, relative to the other.  It was shown in \cite{hide}  that the amplitude of the modulation predicted by the Drude and the plasma prescriptions differ by several orders of magnitude, and not just by a few percent as  with previous setups! In addition to this big improvement, the setup was shown to have a number of distinct advantages that are a
consequence of the {\it differential} character of the measurement scheme.  The first, and perhaps  the most important one, is that the electrostatic calibration becomes totally  unnecessary. This is a consequence of the well known fact that conductors screen out electrostaic fields. Therefore,  electric fields generated by stray electric charges and/or potential patches that may exist either on the corrugated surface of the  exposed Ni surface, or on the {\it flat} exposed surface of the gold over-layer
cannot penetrate inside the gold layer and feel the presence of the hidden Ni corrugation. This implies that  the associated  electrostatic forces  are uncorrelated with the relative phase between the two corrugations, and are thus automatically subtracted out from the modulation.
The second advantage is that the optical properties of Ni or Au are irrelevant too, because  it was shown in \cite{hide} that the modulation of the Casimir force  depends only on the {\it static} magnetic permeability of Ni.  In this way one gets rid of two of the main sources of uncertainty that affect present measurements
of the Casimir force.  It was estimated that with the help of this setup    a clean observation of the controversial thermal Casimir force  should be  possible, with currently available Casimir apparatuses. 

In this paper we present detailed computations for an interesting {\it superconducting} variant of the setup in \cite{hide}. It has been shown in the past that superconductors provide a fertile route to explore thermal effects in Casimir physics \cite{superc,ala1,ala2,super1, super2}. While the superconducting setup described here shares all the distinctive advantages of the previous room-temperature setup,   it dispalys at the same time  certain additional features that make it worth analyzing in detail.   

The plan of the paper is as follows: in Sec. 2 we describe in detail the new setup, and discuss in qualitative terms how it works. In Sec. 3 we present detailed computations of the expected signal based on Lifshitz theory. Finally Sec. 3 presents our conclusions and a discussion of the results.  
 
\section{The setup}
 
\begin{figure}
\includegraphics{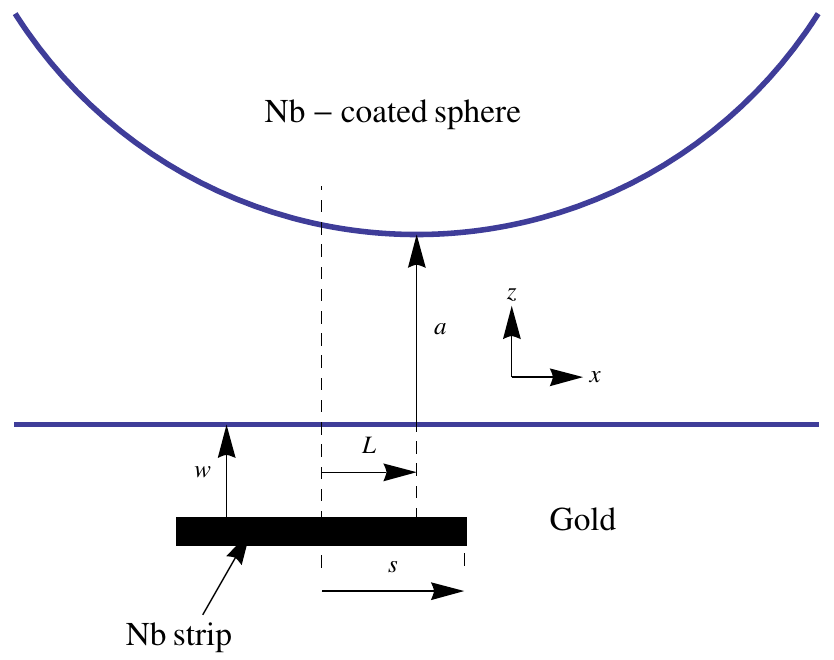}
\caption{\label{setup}  The setup consists of a superconducting Nb-coated sphere above a  superconducting Nb strip (shown in black) included in a gold matrix with a flat exposed surface. }
\end{figure}

The setup, schematically shown in Fig.\ref{setup},  consists of a superconducting Nb sphere of radius $R$ above a thick superconducting Nb strip (shown in black) of half-width $s$. The crucial feature of the setup is that the Nb strip is fully embedded in a gold matrix whose flat exposed surface is parallel to the Nb strip.   
The interesting quantity to consider is  the {\it difference} 
\be\Delta F=F_{\rm Nb}-F_{\rm Au}\ee among  the magnitude $F_{\rm Nb}$ of  the Casimir force that obtains when the tip of the sphere is right above the center of the Nb strip and the magnitude $F_{\rm Au}$ of  the Casimir force that obtains when the tip of the sphere is instead above a point of the gold plate which is far from the Nb strip. 

A little reasoning shows that by measuring  $\Delta F$ it should be easily possible to discriminate which among the Drude and the plasma prescriptions is the correct one. The argument is analogous to that of \cite{hide}, but for the sake of the reader we find it useful to go thru it in some detail.
According to the scattering approach \cite{lambr,rahi}, the Casimir force between the Nb sphere and the Nb-Au plate in Fig.\ref{setup}  can be expressed as a discrete sum over  the so-called imaginary Matsubara  frequencies $\xi_l$ of a series whose terms involve the product of the respective scattering coefficients for em waves of frequency ${\rm i} \,\xi_l$.   The Matsubara modes that contribute significantly are those with frequencies less than a few times the characteristic frequency $\omega_c=c/(2a)$, where $a$ is the minimum sphere-plate separation.    It is easy to convince oneself that  {\it non-vanishing} Matsubara modes give a negligible contribution to $\Delta F$, provided that  the minimum thickness $w$ of the gold matrix  at points above the Nb strip is much larger than    
the characteristic penetration depth $\delta$ of the non-vanishing Matsubara modes in gold (typically $\delta \simeq$ 10 nm or so), a condition that we assume to be satisfied from now on. Indeed, for 
\be w \gg \delta \ee non-vanishing Matsubara modes cannot propagate deep enough inside the gold matrix to reach the Nb strip.  The contribution $F^{(l \neq 0)}$ of these modes to the Casimir force is therefore practically the same, whether the sphere tip is right above the Nb strip or far from it and  we conclude:
\be
\Delta F^{(l \neq 0)} \simeq 0\;.\label{nneq0}
\ee   
Let us consider now the $\omega=0$ modes. They come in two types, i.e. TE and TM $\omega=0$ modes. Since TM $\omega=0$ modes represent static electric fields, they are clearly screened out by the gold overlayer and therefore their contribution to $\Delta F$ is zero:
\be
\Delta F^{({\rm TM};0)} = 0\;.\label{TM0}
\ee
We are left then with the TE $\omega=0$ modes. They represent {\it static magnetic fields}, and whether they contribute or not to $\Delta F$ crucially depends on the adopted prescription. Consider first the Drude prescription.
When modeled as a Drude metal, gold is transparent to static magnetic fields and therefore when the tip of the sphere is far from the Nb strip, the contribution of the  TE $\omega=0$ modes to $F_{\rm Au}$ is zero:
\be
F_{\rm Au}^{({\rm TE};0)}=0\;.\label{TE0NbAu}
\ee  This is not so when the tip of the sphere is above the superconducting Nb strip, because then the TE $\omega=0$ modes can reach unimpeded the Nb strip.  With respect to the TE $\omega=0$ modes   the presence of the Au matrix is indeed immaterial, and everything goes as if  the Nb sphere were facing a "naked" Nb strip. Since superconductors  expel static magnetic fields (Meissner effect) they are strongly scattered by it (we assume that the thickness and width of the Nb strip are both much larger than the penetration depth of magnetic fields).  As a result, within the Drude approach $\Delta F$ receives a significant contribution $F^{({\rm TE};0)}_{\rm Nb}$ from TE $\omega=0$ modes:
\be
\Delta F^{({\rm TE};0)} \simeq F^{({\rm TE};0)}_{\rm Nb}\;,\;\;\;\;({\rm Drude\;prescription}).\label{TE0NbNb}
\ee
In view of Eqs. (\ref{nneq0}-\ref{TE0NbNb}), we arrive at the important conclusion that within the Drude prescription the change $\Delta F$ in the Casimir force originates almost entirely from the contribution to the Casimir force from the TE $\omega=0$ modes, when  the Nb sphere in on top of the Nb strip:
\be
\Delta F \simeq F^{({\rm TE};0)}_{\rm Nb}\;,\;\;\;\;({\rm Drude\;prescription}).\label{drudepre}
\ee
Things go differently with the plasma prescription. While the considerations that led us to Eqs. (\ref{nneq0}-\ref{TM0}) remain unchanged, an important difference arises with regards to the crucial TE $\omega=0$ modes. When modeled as a dissipansionless plasma, gold effectively behaves as a superconductor and it screens out static magnetic fields within a penetration depth $\delta_0=c/\omega_p \simeq 20$ nm. Therefore, for $w \gg \delta_0$ not even the TE $\omega=0$ modes can propagate deep enough in the gold matrix to probe the Nb strip, and similarly to 
the case of the non-vanishing Matsubara modes, we conclude that  within the plasma prescription the TE $\omega=0$ modes also give a negligble contribution to $\Delta F$:
\be
\Delta F^{({\rm TE};0)} \simeq 0\;\label{TE0pl}\;\;\;\;\;\;({\rm plasma\;prescription}).
\ee
In view of  Eqs. (\ref{nneq0}-\ref{TM0}) we conclude
\be
\Delta F \simeq  0\;,\;\;\;\;({\rm plasma\;prescription}).
\ee
The conclusion of the above arguments is that  by measuring $\Delta F$ in this setup,  a clear-cut discrimination  among the two prescriptions should be possible.


 \section{Computing $\Delta F$}

We now turn to the computation of the  Casimir force change $\Delta F$ for our superconducting setup. The framework for computing the Casimir force between real materials  is provided by Lifshitz theory \cite{lifs}. In its original version the material boundaries of the system are planar dielectrics ($\mu=1$)  fully described by the respective frequency-dependent  (complex) dynamical permittivity $\epsilon(\omega)$. The theory was later generalized to deal with layered magneto-dielectric plates in \cite{tomas}.   
In our computations we shall assume, as it usually the case in Casimir experiments, that the sphere radius $R$ is much larger than the tip-plate separation $a$. Under this 
condition it is possible to use the so-called Proximity Force Approximation (PFA) to estimate the Casimir force.  The PFA has been widely used to interpret most  Casimir experiments \cite{book2}  (see  \cite{kruger} for more applications of the Proximity Approximation). 
Recently, it has been shown that the PFA represents  the leading term in a gradient expansion of the Casimir force, in powers of the slopes of the bounding surfaces \cite{fosco2,bimonte3,bimonte4}. The gradient expansion shows that the PFA is asymptotically exact in the zero-curvature limit, and with its help it is now possible  to estimate the error introduced by the PFA.  
Besides $R \gg a$, two further conditions are required to ensure the validity of the PFA in our strip-sphere setup.  To be definite, we let $(x,y,z)$ a cartesian coordinate system such that the $(x,y)$ coordinates span the exposed surface    of the gold plate, placed at $z=0$, while $z=a$ is the plane containing the sphere tip. Consider first the Casimir force $F_{\rm Nb}$ which is observed when the tip of the Nb sphere is placed above the center of the Nb strip. To ensure that we can neglect the effect of the sharp edges of the strip, we shall 
assume that the half-width $s$ of the Nb strip is much larger than the typical radius $\rho=\sqrt{a R}$ of the circular region of the plate that contributes significantly to the Casimir interaction: $s \gg \rho$.  When this condition is fulfilled, the force $F_{\rm Nb}$ is undistinguishable from the force ${\tilde F}_{\rm Nb}$ between the Nb sphere and a plane-parallel two-layers slab consisting of a gold layer of thickness $w$ on top of an infinite Nb planar slab (we recall that the thickness of the Nb strip was assumed to be large compared to the penetration depth of em field in Nb). Consider now the force $F_{\rm Au}$ that is observed when the sphere tip is far from the Nb strip. In practical terms, if the sphere tip is placed above a point of the Au plate whose horizontal distance  from the nearest edge of the Nb strip is much larger than $\rho$, $F_{\rm Au}$ becomes undistinguishable from the Casimir force ${\tilde F}_{\rm Au}$ between the sphere and an infinite homogeneous gold slab. Under these conditions, we can then write for $\Delta F$:
\be
\Delta F \simeq  {\tilde F}_{\rm Nb}-{\tilde F}_{\rm Au} \simeq {F}_{\rm NbAuNb}^{(\rm PFA)}-{F}_{\rm NbAu}^{(\rm PFA)}=2 \pi R ({ {\cal F}}_{\rm NbAuNb}-{ {\cal F}}_{\rm NbAu})\;,\label{forpfa}
\ee 
where ${ { F}}_{\rm NbAuNb}^{(\rm PFA)}$ and ${ { F}}_{\rm NbAu}^{(\rm PFA)}$ denote the PFA expressions for the Casimir force between a Nb sphere and a two-layer Au-Nb planar slab and a gold planar slab, respectively, while  ${ {\cal F}}_{\rm NbAuNb} $ and ${\tilde {\cal F}}_{\rm Au} $ denote the unit-area Casimir free energies  for a plane-parallel system consisting of a Nb planar slab at distance $a$ from a two-layer Au-Nb planar slab and a gold planar slab, respectively, and in the last passage we used the well-known PFA formula for the Casimir free energy of a sphere-plate system:
\be
F^{(\rm PFA)}_{\rm sp-pl} = 2 \pi R {\cal F}.
\ee   
According to  \cite{tomas}, the Casimir free-energy  ${\cal F}$ (per unit area) between two magnetodieletric possibly layered plane-parallel   slabs $S_{j}$, $j=1,2$ at distance $a$ in vacuum  can be represented by the formula: 
$$
{\cal F}(T,a)=\frac{k_B T}{2 \pi}\sum_{l=0}^{\infty}\left(1-\frac{1}{2}\delta_{l0}\right)\int_0^{\infty} d k_{\perp} k_{\perp}  
$$
\be
\times \; \sum_{\alpha={\rm TE,TM}} \log \left[1- {e^{-2 d q_l}}{R^{(1)}_{\alpha}({\rm i} \xi_l,k_{\perp})\;R^{(2)}_{\alpha}({\rm i} \xi_l,k_{\perp})} \right]\;,\label{lifs}
\ee
where $k_B$ is the Boltzmann constant, $\xi_l=2 \pi l k_B T/\hbar$ are the (imaginary) Matsubara frequencies, $k_{\perp}$ is the modulus of the in-plane wave-vector, $q_l=\sqrt{\xi_l^2/c^2+k_{\perp}^2}$, and $R^{(j)}_{\alpha}({\rm i} \xi_l,k_{\perp})$ is the reflection coefficient of slab $j$ for polarization $\alpha$.  The Casimir free energy ${\cal F}_{\rm NbAuNb}$
for a planar Nb-Au-Nb system, can be  obtained from  Eq. (\ref{lifs}) by substituting  $R^{(1)}_{\alpha}$ by the Fresnel reflection coefficient $r_{\alpha}^{(0{\rm Nb})}$ of a Nb slab (given in Eqs.(\ref{freTE}) and (\ref{freTM}) below, with $a=0$, $b=$Nb),  and $R^{(2)}_{\alpha}$ by the reflection coefficient  $R_{\alpha}^{(0{\rm AuNb})}$ of a two-layer planar slab consisting of a layer of thickness $w$ of gold on a Nb substrate.  The latter reflection coefficient has the expression:
\be  
R_{\alpha}^{(0{\rm Au Nb})}({\rm i} \xi_l,k_{\perp};w)=\frac{r_{\alpha}^{(0{\rm Au})}+e^{-2\,w\, k_l^{({\rm Au})}}\,r_{\alpha}^{({\rm Au Nb})}}{1+e^{-2\,w\, k_l^{({\rm Au})}}\,r_{\alpha}^{(0{\rm Au})}\,r_{\alpha}^{({\rm Au Nb})}}\;.
\ee   
Here $r^{(ab)}_{\alpha}$ are the Fresnel reflection coefficients for a planar dielectric (we set $\mu=1$ for all materials) interface separating medium a from medium b: 
\be
r^{(ab)}_{\rm TE}=\frac{  \,k_l^{(a)}-  \,k_l^{(b)}}{ k_l^{(a)}+ \,k_l^{(b)}}\;,\label{freTE}
\ee
\be
r^{(ab)}_{\rm TM}=\frac{\epsilon_{b} ({\rm i} \xi_l) \,k_l^{(a)}-\epsilon_{a}({\rm i} \xi_l) \,k_l^{(b)}}{\epsilon_{b}({\rm i} \xi_l) \,k_l^{(a)}+\epsilon_{a}({\rm i} \xi_l) \,k_l^{(b)}}\;,\label{freTM}
\ee
where
$ k_l^{(a)}=\sqrt{\epsilon_a({\rm i} \xi_l)  \xi_l^2/c^2+k_{\perp}^2}\;$, 
 $\epsilon_a$  denotes the electric   permittivity of medium $a$, and we define $\epsilon_0=1$. 

The values $\epsilon_{\rm Au}({\rm i} \xi_l)$ of the permittivity of gold were computed using a six-oscillator model of the form:
\be
\epsilon({\rm i } \xi)=1+\frac{\omega_{\rm Au}^2}{\xi (\xi+\gamma_{\rm Au})}+\sum_{j=1}^{6} \frac{g_j}{\xi^2+\omega_j^2+\gamma_j \xi}\;.
\ee
The first two terms provide the familiar Drude formula and account for the contribution of conduction electrons, while the oscillator terms describe core electrons. For the oscillator parameters $g_j$, $\omega_j$ and $\gamma_j$, which are expected to be temperature independent, we used their room temperature values \cite{decca2}. For the plasma frequency, which is also temperature independent, we used the standard value $\omega_{\rm Au}=9 \,{\rm eV}/\hbar$.   The  relaxation parameter $\gamma_{\rm Au}$ is  temperature-dependent, but for cryogenic temperatures it   reaches a saturation value $\gamma_{\rm Au}^{(\rm s)}$ that depends on the amount of impurities and on the fabrication process. In our computations we thus set $\gamma_{\rm Au}=\gamma_{\rm Au}^{(\rm s)}$. It is customary to express $\gamma_{\rm Au}^{(\rm s)}$ in terms of the room temperature relaxation parameter $\gamma_{\rm Au}|_{\rm room}$ ($\gamma_{\rm Au}|_{\rm room}=0.035$ eV/$\hbar$) as:     $ \gamma_{\rm Au}^{(\rm s)}=\gamma_{\rm Au}|_{\rm room}/ RRR_{\rm Au}$, where $RRR_{\rm Au}$ is the so-called residual-resitance ratio. 

The values of $\epsilon_{\rm Nb}({\rm i} \xi_l)$ for superconducting Nb were  computed using the {\it local} limit \cite{zimm,berl} of the Mattis-Bardeen \cite{mattis} analytic formula for the conductivity of a BCS superconductor.  The analytic continuation of this formula to the imaginary frequency axis was worked out in \cite{super2}. We remark that the local approximation
is expected to be reliable for separations $a$ that are sufficiently large compared to the superconducting coherence lenght $\xi$ (typically for separations $a \gtrsim$ 50 nm \cite{ala1,ala2,raul}), in which range the effect of spatial dispersion can be safely neglected. While we address the reader to Ref. \cite{super2} for a detailed discussion of the Mattis-Bardeen dielectric function, we briefly recall here its main features. The permittivity $\epsilon_{\rm Nb}({\rm i} \xi)$ depends parametrically on the temperature, on the plasma and relaxation frequencies of Nb, $\omega_{\rm Nb}$ and $\gamma_{\rm Nb}$ and on the (temperature-dependent) BCS gap $\Delta(T) $ (in our computations we used the approximate expression of the BCS gap function $\Delta(T)$ given in \cite{gap1,gap2} ). As the temperature $T$ is increased towards the critical temperature $T_c$ ($T_c=9.25$ K), the BCS gap $\Delta(T)$ approaches zero and the Mattis-Bardeen formula approaches the Drude formula for the permittivty of normal metals. For the plasma frequency $\omega_{\rm Nb}$ we took $\omega_{\rm Nb}=9.3$ eV/$\hbar$. For  the relaxation parameter $\gamma_{\rm Nb}$ we took again  its saturation value $\gamma_{\rm Nb}=\gamma_{\rm Nb}^{(\rm s)}=\gamma_{\rm Nb}|_{\rm room}/ RRR_{\rm Nb}$. For simplicity
in our computations we used the realistic common value $RRR_{\rm Au}=RRR_{\rm Nb}=RRR=5$ both for Au and Nb.

\begin{figure}
\includegraphics{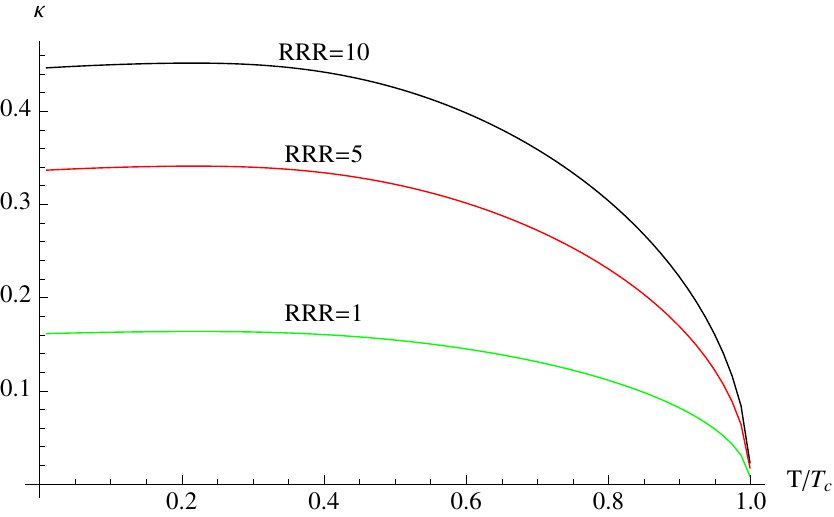}
\caption{\label{fig2}  The parameter $\kappa$ versus $T/T_c$ for Nb }
\end{figure}
The most important feature of the Mattis-Bardeen formula for our purposes is its double pole at zero frequency:
\be
\epsilon_{\rm Nb}({\rm i} \xi_l) \sim \frac{\omega_s^2(T)}{\xi^2} + {\rm O}\left( \frac{\log (\Delta/\xi)}{\xi}\right)\;,\;\;\;\;{\rm for}\;\xi \rightarrow 0\;,
\ee  
where $\omega_s^2(T)=\kappa\, \omega_{\rm Nb}^2$.
The coefficient $\kappa$, which is positive and always less than one, is a function of the {\it impurity} parameter $\eta=\gamma_{\rm Nb}/(2 \Delta)$ and of the {\it reduced} temperature $t=k_B T/(2 \Delta)$. In the limit  $\Delta \rightarrow 0$ (i.e. for $T \rightarrow T_c$) $\kappa$ vanishes, as it should because for vanishing gap the Mattis-Bardeen formula approaches the Drude formula. The coefficient $\kappa$  vanishes also in the impure limit $\eta \rightarrow \infty$, while in the opposite pure limit ($\eta \rightarrow 0$) $\kappa$ approaches one. In Fig. \ref{fig2} we plot the parameter $\kappa$ versus $T/T_c$ for Nb, for three values of RRR.
     
We can now present the results of our numerical computations for $\Delta F$. It is useful to separate the contribution $\Delta F^{({\rm TE};0)}$ of the TE $\omega=0$   mode   from the combined contributions $\overline{\Delta { F}}=  \Delta F^{({\rm TM};0)}+ \Delta F^{(l \neq 0) }$  of the TM $\omega=0$ plus all the non-vanishing  Matsubara modes:
$
\Delta F=\Delta F_{TE}^0+\overline{\Delta { F}}\;.
$
\begin{figure}
\includegraphics{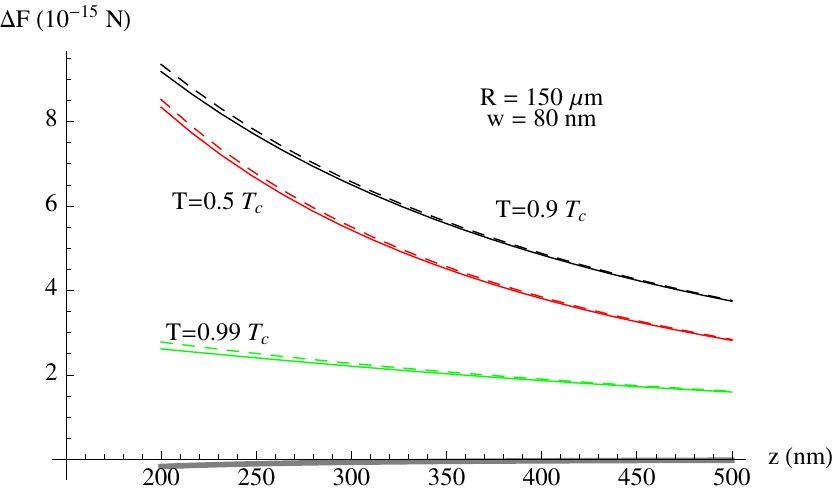}
\caption{\label{fig3}   $\Delta F$ (in units of $10^{-15}$ N) versus sphere-plate separation (in nm) for $w=80$ nm, and $RRR=5$. The solid black, red and green lines represent the Drude prediction for $T/T_c=0.9$ (black curve), $T/T_c=0.5$ (red curve)  and for $T/T_c=0.99$ (green line). The dashed lines represent the respective  contributions of the TE $\omega=0$ mode alone. The thick grey line represents the plasma-model prediction.}
\end{figure}
In Fig. \ref{fig3} we plot $\Delta F$   (in units of $10^{-15}$ N) versus separation $a$ (in nm) for $R=150\; \mu$m and a minimum thickness $w=80$ nm of the gold overlayer, for three different temperatures $T=0.9 T_c$ (black curve), $T=0.5 T_c$ (red curve) and for $T=0.99 T_c$ (green curve).  The three dashed lines represent the respective contributions  $\Delta F^{({\rm TE};0)}$ of the TE  $\omega=0$ mode alone.  All these lines were computed using the Drude prescription. The thick grey line represents the modulation that obtains if the plasma prescription is used. In Fig. \ref{fig4} we plot $\Delta F$   (in units of $10^{-15}$ N) versus $T/T_c$, for $a=300$ nm (all other parameters as before). The blue and thick grey lines represent the Drude and plasma predictions, respectively. The dashed blue line represents the Drude contribution of the TE $\omega=0$ mode alone.
\begin{figure}
\includegraphics{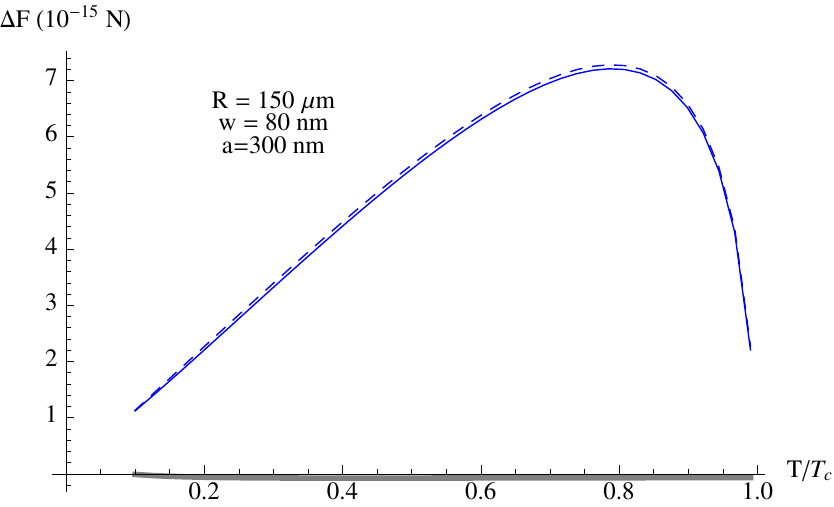}
\caption{\label{fig4}   $\Delta F$ (in units of $10^{-15}$ N) versus $T/T_c$ for $a=300$ nm, $w=80$ nm, and $RRR=5$.  The blue lines represent the Drude prediction. The dashed blue line represents the Drude  contribution of the TE $\omega=0$ mode alone. The thick grey line represents the plasma-model prediction.}
\end{figure}

We see from Figs. \ref{fig3} and \ref{fig4}  that the qualitative arguments outlined in Sec. 2 are fully confirmed by our computations. The most important conclusion that can be drawn from Figs.  \ref{fig3} and \ref{fig4} is that the Drude and the plasma prescriptions predict {\it widely} different magnitudes for the
change of the Casimir force. For example, for $T/T_c=0.8$, $a=300$ nm and $w=80$ nm we find that the magnitude of $\Delta F$ predicted by the Drude approach is over two orders of magnitudes larger than the one predicted by the plasma prescription. Such a wide difference should make it possible to easily discriminate between the two prescriptions, provided that the sensitivity of the apparatus is sufficient to detect force differences of the order of $10^{-15} N$.   

Let us consider more closely the Drude prediction for $\Delta F$. As it was argued in Sec.2,   the Drude model predicts a temperature dependent change of the Casimir force $\Delta F$, which is almost entirely determined by the TE $\omega=0$ mode.  From Eq. (\ref{drudepre}), and using the same arguments that led us to Eq. (\ref{forpfa}) we estimate:
$$
\Delta F^{(\rm Dr)} \simeq 2 \pi R {\cal F}^{(\rm TE;0)}_{\rm NbAuNb}=\frac{k_B T R}{2 a^2}  \times   
$$
\be
\times\;\int_0^{\infty}\!\!\! d y y
 \; \log \left[1-e^{-2 y}\left(\frac{y-\sqrt{y^2+\omega_s^2(T) a^2/c^2}}{y+\sqrt{y^2+\omega_s^2(T) a^2/c^2}}\right)^2 \right]\;.
\ee
We see that within the Drude prescription the dimensionless change of the Casimir force  $a^2 \Delta F^{(\rm Dr)}/(k_B T R)$  is entirely determined by the dimensionless parameter $\omega_s^2(T) a^2/c^2$,
and   it is otherwise independent of the optical properties of the materials. This feature makes the Drude prediction for $\Delta F$ robust. 

The much smaller plasma prediction for $\Delta F$ receives a significant contribution from several Matsubara frequencies, and its precise magntude depends on the detailed optical properties of Au and Nb in the infrared region of the spectrum. 


\section{Conclusions}

We have studied a  superconducting Casimir setup consisting of a Nb sphere and  a Nb strip. The key feature of the device is that the Nb strip is fully immersed into a flat gold matrix, in such a way that the Nb strip is optically hidden by a gold over-layer. The proposed experimental scheme  is based on a differential measurement of the Casimir force, at a point of the gold plate above the Nb strip and away from it.    We have  shown that the Drude and the plasma prescriptions lead to widely different predictions for the change of the Casimir force. We thus expect that by this setup  it should  be possible to easily discriminate between them, provided that the sensitivity of the apparatus is of the order of $10^{-15} N$. 
The superconducting setup discussed here shares the advantages of an analogous setup based on magnetic materials, that was recently proposed by the author \cite{hide}. In particular  it requires neither an electrostatic calibration of the apparatus, nor any detailed 
knowledge of the optical properties of the involved materials.

\end{document}